\documentclass{Interspeech}

\interspeechcameraready
\usepackage{placeins} % This provides \FloatBarrier, which can be used to make images appear where you want, or at least not too far away.

\usepackage{cite}  % this package compresses references, e.g., [1-3] instead of [1, 2, 3]
\usepackage{multirow}
\usepackage{tablefootnote}

\graphicspath{ {./figures/} }

\title{Investigating Stochastic Methods for Prosody Modeling in Speech Synthesis}  % Let It Flow: Prosody Prediction via Stochastic Generative Models

\author[affiliation={1}, equalcontribution]{Paul}{Mayer}
\author[affiliation={2}, equalcontribution]{Florian}{Lux}
\author[affiliation={2}]{Alejandro}{Pérez-González-de-Martos}
\author[affiliation={2}]{Angelina}{Elizarova}
\author[affiliation={1}]{\\Lindsey}{Vanderlyn}
\author[affiliation={1}]{Dirk}{Väth}
\author[affiliation={1}]{Ngoc Thang}{Vu}

\affiliation{}{University of Stuttgart}{Germany}
\affiliation{}{AppTek GmbH}{Germany}

\email{flux@apptek.com}


\begin{document}

\maketitle

\begin{abstract}
% 1000 characters. ASCII characters only. No citations.
While generative methods have progressed rapidly in recent years, generating expressive prosody for an utterance remains a challenging task in text-to-speech synthesis. This is particularly true for systems that model prosody explicitly through parameters such as pitch, energy, and duration, which is commonly done for the sake of interpretability and controllability. In this work, we investigate the effectiveness of stochastic methods for this task, including Normalizing Flows, Conditional Flow Matching, and Rectified Flows.
We compare these methods to a traditional deterministic baseline, as well as to real human realizations. Our extensive subjective and objective evaluations demonstrate that stochastic methods produce natural prosody on par with human speakers by capturing the variability inherent in human speech. Further, they open up additional controllability options by allowing the sampling temperature to be tuned. 
\end{abstract}

\section{Introduction}

In recent years, text-to-speech (TTS) systems have made significant advancements in expressivity and naturalness, closing the gap to human speech in multiple instances \cite{tan2024naturalspeech, liu2022delightfultts, shen2023naturalspeech, liu2021delightfultts}. However, there remains an inherent trade-off between controllability and expressivity. While modern language modeling approaches to TTS have demonstrated impressive performance in generating human-like speech \cite{Betker2023, wang2023neuralcodeclanguagemodels, chen2023vector}, they operate as a black box without providing mechanisms to influence the prosody of an utterance on a fine-grained level. This constraint poses challenges in applications that require precise control over the generated speech, such as speaker anonymization and voice privacy \cite{meyer2023prosody}, literary studies and digital humanities \cite{PoeticTTS}, adversarial defense against deepfakes \cite{wang2025asvspoof}, or content generation and high-quality dubbing \cite{xie2024towards}. 

Explicit modeling of phoneme durations, as well as pitch and energy contours \cite{ren2019fastspeech, ren2020fastspeech, lancucki2021fastpitch}, allows for human intervention in the generation pipeline to refine the speech output. However, these systems often lack expressivity, as their prosody prediction mechanisms struggle to capture the variability present in natural speech. This limitation results in synthesized speech that, while controllable, may sound less dynamic and varied compared to approaches that model prosodic parameters like duration implicitly, for example through autoregression. 

As mentioned previously, however, systems that utilize explicit prosody prediction are still a necessity for many important applications. Hence, we address this bottleneck in this work by investigating the use of state-of-the-art generative methods, such as Normalizing Flows (NF) \cite{rezende2015variational}, Conditional Flow Matching (CFM) \cite{lipman2023flowmatchinggenerativemodeling}, and Rectified Flows (RF) \cite{Liu2022FlowSA, liu2022rectifiedflowmarginalpreserving} for explicit prosody prediction. 

One of the earliest works introducing advances towards this goal is the VITS architecture \cite{kim2021conditional}, which explores the idea of using an NF module to predict durations in a stochastic fashion. In their evaluation, they confirm the improved ability of their stochastic duration predictor to adapt to various speaking rhythms. VarianceFlow \cite{lee2022varianceflow} takes this NF module and extends its application to pitch and energy prediction. Finally, and most recently, \cite{mehta2024should} confirm the benefits of stochastic methods for duration modeling using a CFM module. They show in their evaluation that especially for conversational scenarios, speech produced from a system with stochastic duration prediction outperforms deterministic systems by a significant margin. They do, however, find no benefit in using the CFM module over a deterministic module for pitch prediction.

\begin{figure}[t]
    \centering
    \includegraphics[width=\linewidth]{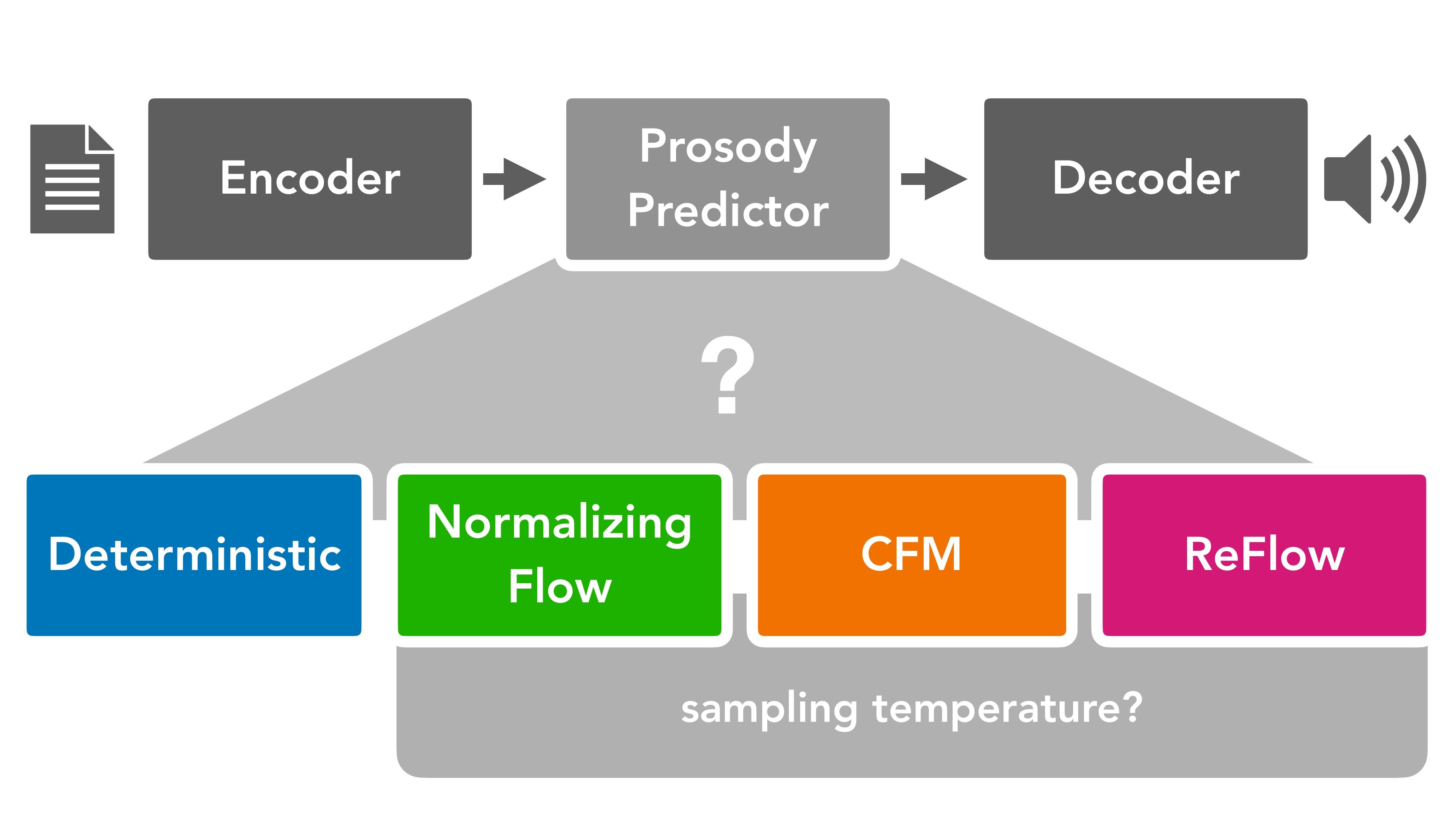} % TODO see if 1.0 also fits after everything is done
    \caption{An overview of the questions we want to answer: Which method is most suitable to model prosody in speech synthesis and which sampling temperature works best?}
    \label{fig:teaser}
\end{figure}

Extending this relatively underexplored line of research, we conduct a thorough investigation of modeling pitch, energy, and durations using NF, CFM, RF, and a deterministic baseline, as visualized in Figure \ref{fig:teaser}. We also include human samples as an upper bound in the comparisons. We perform objective evaluations to explore whether prosodic parameters should be modeled jointly or separately, the impact of their order, as well as the effects of the sampling temperature and the generative method used. Furthermore, we conduct a study with human raters to compare systems and temperatures in terms of prosodic naturalness and prosodic diversity.

We find that 1) duration prediction benefits from modeling prosodic parameters as cascade instead of jointly, 2) the impact of the order of pitch and energy prediction is negligible, and 3) the variance in generated speech can be controlled effectively using the sampling temperature. Human judgment further reveals that 4) naturalness and diversity have an inverse relationship, even for human recordings, and 5) the RF based model offers the most favorable trade-off between the two desired properties. Additionally, we provide open-source code\footnote{\url{https://github.com/DigitalPhonetics/IMS-Toucan/tree/StochasticProsodyModeling}} to support reproducibility and foster the adoption of these techniques.

\section{Methods} 
\subsection{Overall Pipeline}
The pipeline follows the architecture of ToucanTTS \cite{lux2023toucan, lux2024massive} due to its modularity and open-source implementation. It builds on FastSpeech 2 \cite{ren2019fastspeech, ren2020fastspeech} and incorporates several key improvements, such as the use of articulatory features as the input representation \cite{lux2022laml} and a Conformer-based encoder and decoder \cite{gulati2020conformer}. Similar to FastPitch \cite{lancucki2021fastpitch}, pitch and energy values are averaged over the duration of each phoneme and normalized on an utterance level to make the contours speaker-independent \cite{lux2023exact}. Speaker characteristics are conditioned on a pre-trained ECAPA-TDNN \cite{ecapa} speaker embedding model provided by Speechbrain \cite{speechbrain}. Additionally, the CFM PostNet, adapted from MatchaTTS \cite{mehta2024matcha}, refines the generated spectrogram. While the training of this CFM module follows a stochastic process, %which allows to account for variability in the production of the spectrogram, 
the sampling temperature is set to 0.0 during inference, to ensure that the only variance in the speech signal stems from the variance predictors which are conditioned on the speaker and the encoder output. The predictors estimate pitch, energy, and duration for each phoneme. During training, ground-truth values are available and teacher forcing is employed to prevent cascading errors. The following sections take a closer look at the structure and functionality of the different variance predictors that this work focuses on.

\subsection{Order of Predictors}
The first setup we consider uses the predictors sequentially with residual connections. This cascading approach allows each predictor to be conditioned on previously estimated prosodic features, potentially capturing hierarchical dependencies between them. This also raises the question of which order these should be predicted in. Since both the pitch contour and the energy contour add information to the latent space, their order could have an impact. This is not the case for the duration predictor, as it does not modify the latent space since it is purely used for upsampling. Hence, we always predict duration last and only explore the order of the pitch and energy predictors in the cascade. An entirely different approach  % which we also explore, 
is to use a single predictor with a three-dimensional output, where pitch, energy, and duration are predicted simultaneously from the same module.  
 
\subsection{Deterministic Baseline}
The baseline system adopts a deterministic approach inspired by the convolutional variance predictors in FastSpeech 2 \cite{ren2020fastspeech}. The encoder output is transformed into a one-dimensional sequence in a purely deterministic way, which collapses all valid prosodic realizations of an utterance into a single averaged most-likely realization. During inference, the same sentence will always be produced in exactly the same way. While this approach leads to very robust and consistent realizations, it may not be suitable for all use-cases, since the most-likely realization is not always the desired one.

\subsection{Probabilistic Predictors} 
While the deterministic model faces the inherent one-to-many problem of TTS, where an individual point in the input can be mapped to multiple valid targets, the stochastic models solve this elegantly by treating this task as a transport problem between distributions. The input distribution is a simple Gaussian for all our models, which they transform into the distribution of valid prosodic realizations given the encoded phoneme sequence as condition using different techniques. 

\medskip\noindent\textbf{Normalizing Flow} \cite{rezende2015variational} The first of these techniques models the flow from the data distribution to the noise distribution using a series of invertible transformations. During inference, we use the inverse transformation to map the noise to the data. Since it is comparably easy to learn to predict noise from data, yet the inverse is mathematically tractable, we can learn a flow between the two distributions in a fast and stable way. This approach works in a non-autoregressive fashion and requires only a single step. However, the transformation is still gradual through the use of many blocks of these invertible transformations. A major issue with this approach, however, is that the couplings between noise and data are arbitrary, which causes NFs to be unable to fully capture complex distributions.

\medskip\noindent\textbf{Conditional Flow Matching} \cite{lipman2023flowmatchinggenerativemodeling} The next technique is also a type of flow that is defined as a transport function between arbitrary couplings of noise and data. However, this transport function is not realized as an invertible bijective mapping. Instead, it represents a time-dependent vector field $u_t$, which is trained to approximate an existing mathematically tractable transport function. It is used to generate a probability path $p_t$ from $p_0$ to $p_1 \approx q$, where $q$ is the target distribution. While the CFM approach is very general, and can theoretically approximate any transport function between any distributions, it makes sense to use \textit{optimal transport} as the function to be approximated, since this makes the flow move data points along the shortest possible path. During inference, we apply time-discretization to move the points along $p_t$ using an ordinary differential equation (ODE) solver in multiple steps to reduce the deviation that comes with the use of an imperfect ODE solver, such as the Euler solver. The objective function of the CFM model is then 
\begin{equation}    \label{eq:lcfm}
L_{CFM}(\theta) = \mathbb{E}_{t,q(x_1),p_t(x|x_1)} \lVert u_t(x|x_1)-v_t(x;\theta) \rVert ^ 2
\end{equation}

where $v_t(x,\theta)$ is our neural network with parameters $\theta$ and $x$ representing an observation in the data space $q$. $x_1$ is a condition that is sampled from the noise distribution in the first time step. The flow matching procedure needs to be conditional to make the problem tractable, but it again raises the issue of using an arbitrary coupling, causing crossing paths $p_t$ which prevent the model from learning perfectly straight optimal transport paths.

\medskip\noindent\textbf{Rectified Flow} \cite{Liu2022FlowSA} The final technique we consider is the same as the previous CFM model, however, it turns the arbitrary coupling into a deterministic coupling using a post-training procedure called ReFlow. After the CFM has finished training, we sample $x_1$ from a Gaussian and then use a frozen copy of $p_t$ to solve for $x_{det}$, using many ODE steps $t$ to approximate an exact solution. We then continue training the CFM model on the deterministic coupling of $(x_1, x_{det})$, which straightens $p_t$, aligning it more closely with optimal transport. %Importantly, this does not affect the frozen copy of $p_t$, which we use as teacher in this case.

\section{Experiments}

\begin{table}[t]
    \caption{JS divergence ($\downarrow$) between human and synthetic distributions for various configurations and prosodic variables.}
    \centering
    \begin{tabular}{lccc}
        \toprule
        Model & Pitch JS & Energy JS & Duration JS \\
        \midrule
        Pitch First & 0.7019 & 0.6887 & \textbf{0.4808} \\
        Energy First & 0.7050 & 0.6894 & \textbf{0.4867} \\
        Joint Prediction & 0.7070 & 0.6855 & 0.5621 \\
        \bottomrule
    \end{tabular}
    \label{tab:comparison}
\end{table}

\subsection{Datasets}
\textbf{Training Data:} For this work, we constrain ourselves to read speech, leaving experiments on conversational speech and other more challenging scenarios for future work. We use LibriTTS \cite{zen2019libritts} for its clean and consistent data. This dataset is comprised exclusively of read speech in English and features 2,456 speakers. Since the original purpose of the recordings was audiobooks, the utterances contain more prosodic variation than other types of read speech, facilitating the learning of expressive prosody by our variance predictors. 

\medskip\noindent\textbf{Evaluation Data:} For objective evaluation, we use the RAVDESS dataset \cite{Livingstone2018}. It consists of recordings from 24 professional actors vocalizing two sentences with multiple emotions and intensities. This diversity in expressions for the same sentence allows us to estimate a distribution of human realizations for each sentence.
For subjective evaluation, we utilize the topical emphasis portion of the ADEPT dataset \cite{torresquintero21adept}, which provides three samples per speaker with varying emphasis placement. This choice is motivated by the fact that the realizations of topical emphasis are similarly diverse to the emotional realizations from RAVDESS in terms of prosody, however, they are less salient than the emotions, which facilitates the unbiased diversity estimation by human raters.

\subsection{Setup}
To ensure comparability between the generative methods, all stochastic models use the same hyperparameters. Since we only need to model a one-dimensional sequence in the cascading case, we keep the parameter count very low and in line with the parameter count of the deterministic model. The noise vector has a dimensionality of 8, and we use 6  diffusion transformer layers \cite{peebles2023scalable}, with a kernel size of 5 and a hidden size of 8. For the combined approach, the hyperparameters are scaled up such that the amount of parameters in the model stays constant. The models were trained for 100k steps with a batch size of 32, which allowed all models to converge. The RF model was trained for an additional 10k steps with the ReFlow stage.
This does not require training data, so all models saw the same amount of data. During inference, we use the Euler solver for both ODE based methods with 12 timesteps.

\subsection{Objective Evaluation}
First, we want to explore the impact of order on the performance of the variance predictors. To objectively measure this, we calculate a distributional distance measure between generated samples and human speech. For this, we first estimate the probability density functions (PDFs) using Gaussian kernel density estimation (KDE). We then calculate the Jensen–Shannon (JS) divergence \cite{lin1991divergence, kullback1951information} between the estimated PDFs of the human samples and the synthetic samples. Human references are taken from the RAVDESS dataset and compared to synthetic samples generated using a CFM based predictor following \cite{mehta2024should} with the same sentences and speakers. We see from the results in Table \ref{tab:comparison} that the cascading approaches outperform the joint modeling approach by a large margin for duration prediction. For pitch and energy prediction, the difference is not significant. %, which indicates that it is mostly helpful for the duration prediction to have prior information about the pitch and energy. 
For the order of the predictors within the cascade, the difference is never significant. We can hence conclude that only duration prediction requires prior knowledge of other prosodic variables. We verified these results in informal listening tests and use a cascade of energy$\rightarrow$pitch$\rightarrow$duration for all following experiments. 

\begin{figure}[t] 
    \centering
    \includegraphics[width=\linewidth]{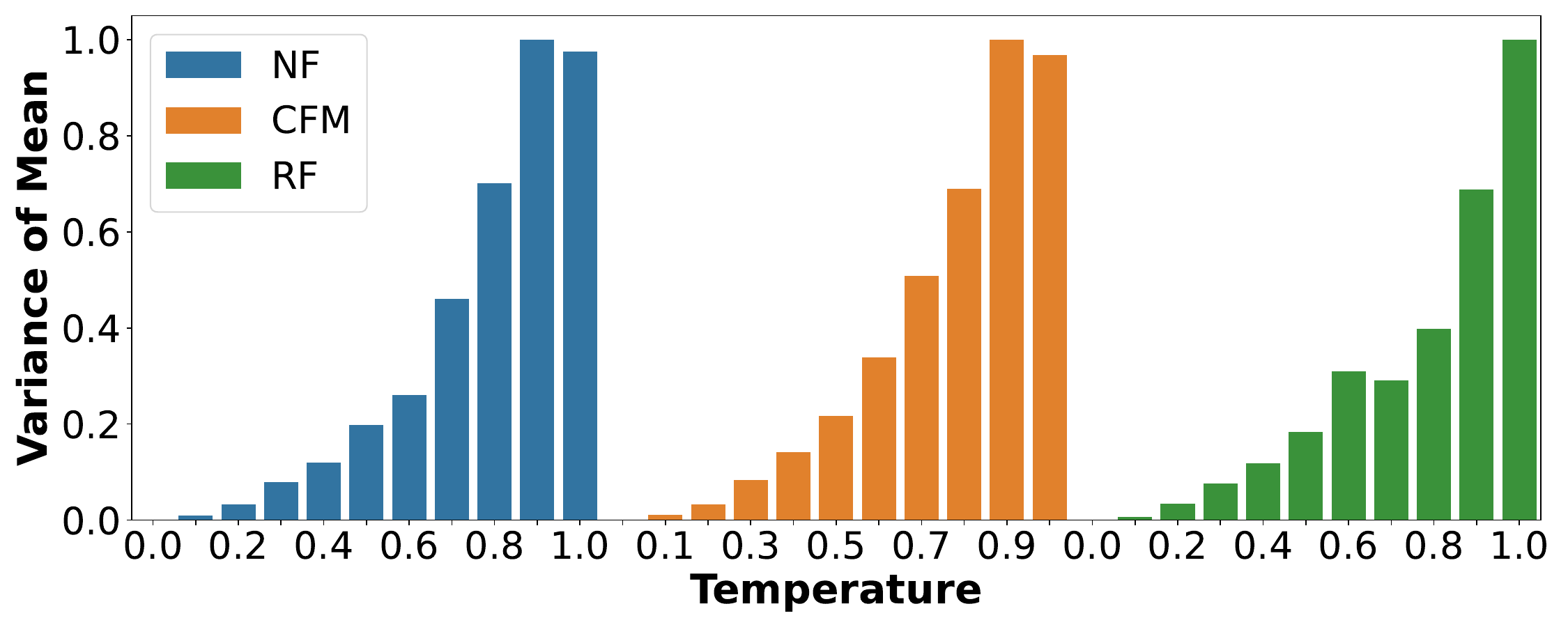}
    \caption{Overview of how the sampling temperature affects the average variance within the \textbf{pitch contour} of an utterance for each system on average.}
    \label{fig:temps_pitch}
\end{figure}

\begin{figure}[t]
    \centering
    \includegraphics[width=\linewidth]{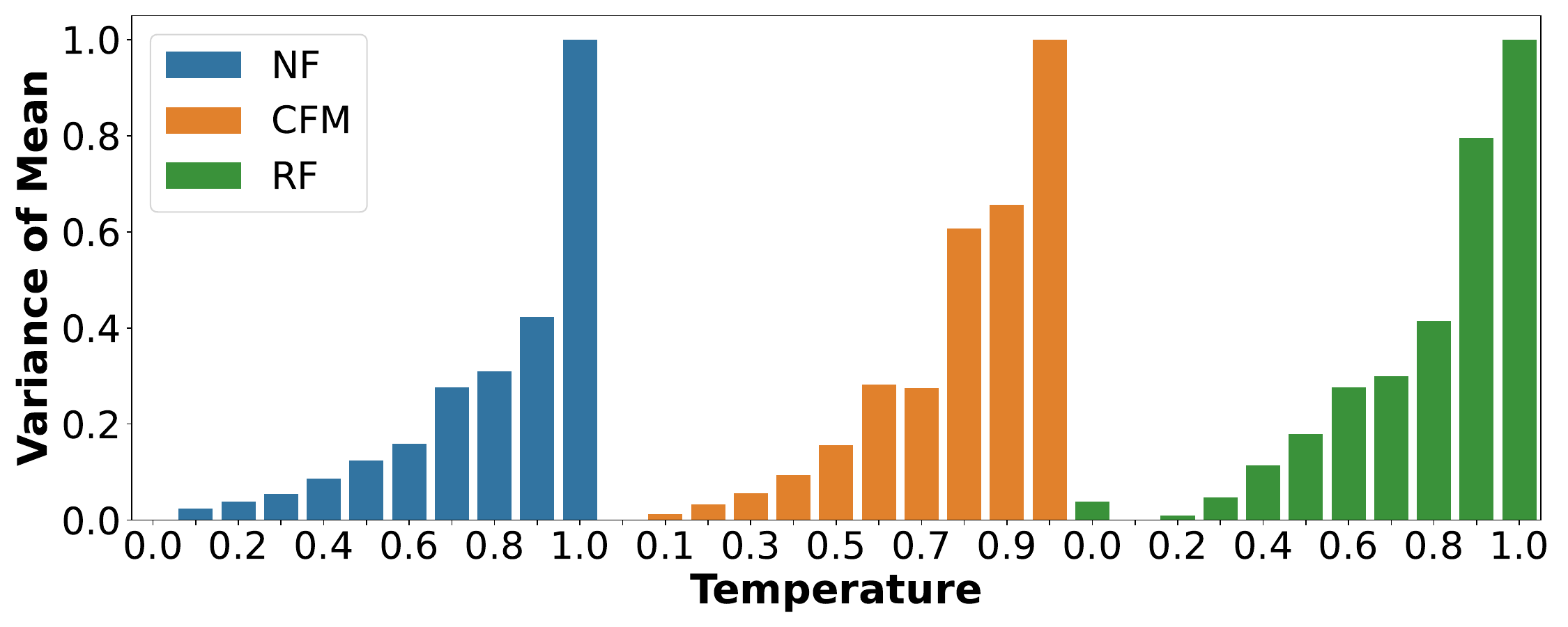}
    \caption{Overview of how the sampling temperature affects the average variance within the \textbf{durations} of an utterance for each system on average.}
    \label{fig:temps_dur}
\end{figure}

To explore whether the sampling temperature can effectively steer the diversity of the predicted prosody, we measure the variance of the prosodic variables over an utterance. In Figure \ref{fig:temps_pitch} and Figure \ref{fig:temps_dur}, we visualize the variance of the average pitch and the variance of the average duration, respectively, across 200 utterances for different temperatures. We can see that the sampling temperature can effectively control the variance for all stochastic methods. The trend, however, does not follow a strictly linear correlation, as the differences become larger for temperatures close to 1.0. Values above 1.0 follow the same trend, however, to such an extreme that we found this not to be a reasonable choice through informal listening tests and hence excluded these higher temperatures from all experiments. An interesting finding of this experiment is that, due to the nearly exponential relation between variance and temperature, a logarithmic transformation on the temperature axis can turn the increase in variance into a linear relation, which may be more intuitive for users of a system that exposes this control mechanism.

\begin{figure}[t] 
    \centering
    \includegraphics[width=\linewidth]{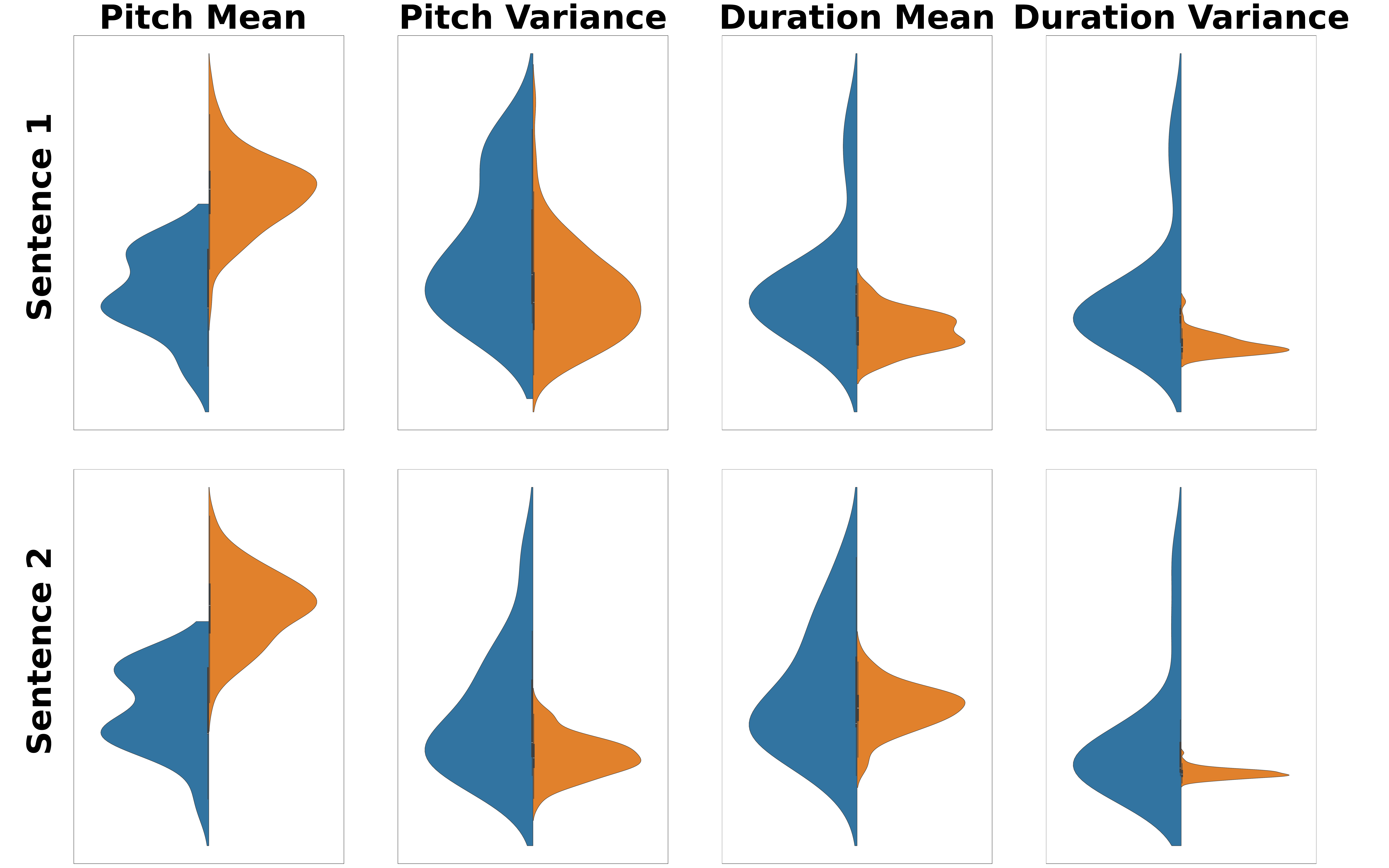}
    \caption{Blue is the human, orange is the CFM system. The distributions are the KDE results of the mean values and the variance values for pitch and duration across different realizations of the same sentence. Each row is based on one sentence.}
    \label{fig:violin}
\end{figure}

We also perform qualitative analysis on individual sentences by visualizing the distributions of means and variances of values across multiple realizations of that sentence. An example of this is shown in Figure \ref{fig:violin}, where the synthetic distribution is generated with the RF based model using a temperature of 0.8. We see that the distribution derived from the human samples (blue, left) is usually very wide and sometimes even has two modes. For the synthetic samples (orange, right), we see that most are unimodal, where the mode generally aligns with the main mode of the human distribution, but is more narrow. This indicates that the full diversity of the human speech is not yet covered by this model, despite trends towards this being visible. 
% It is possible that this could be fixed by making the predictor more powerful through additional layers, which remains to be explored in future work.
We hypothesize that increasing the number of parameters in the prosody prediction modules could lead to larger prosodic diversity, which remains to be explored in future work.

\subsection{Subjective Evaluation}
\begin{figure}[t]
    \centering
    \includegraphics[width=\linewidth]{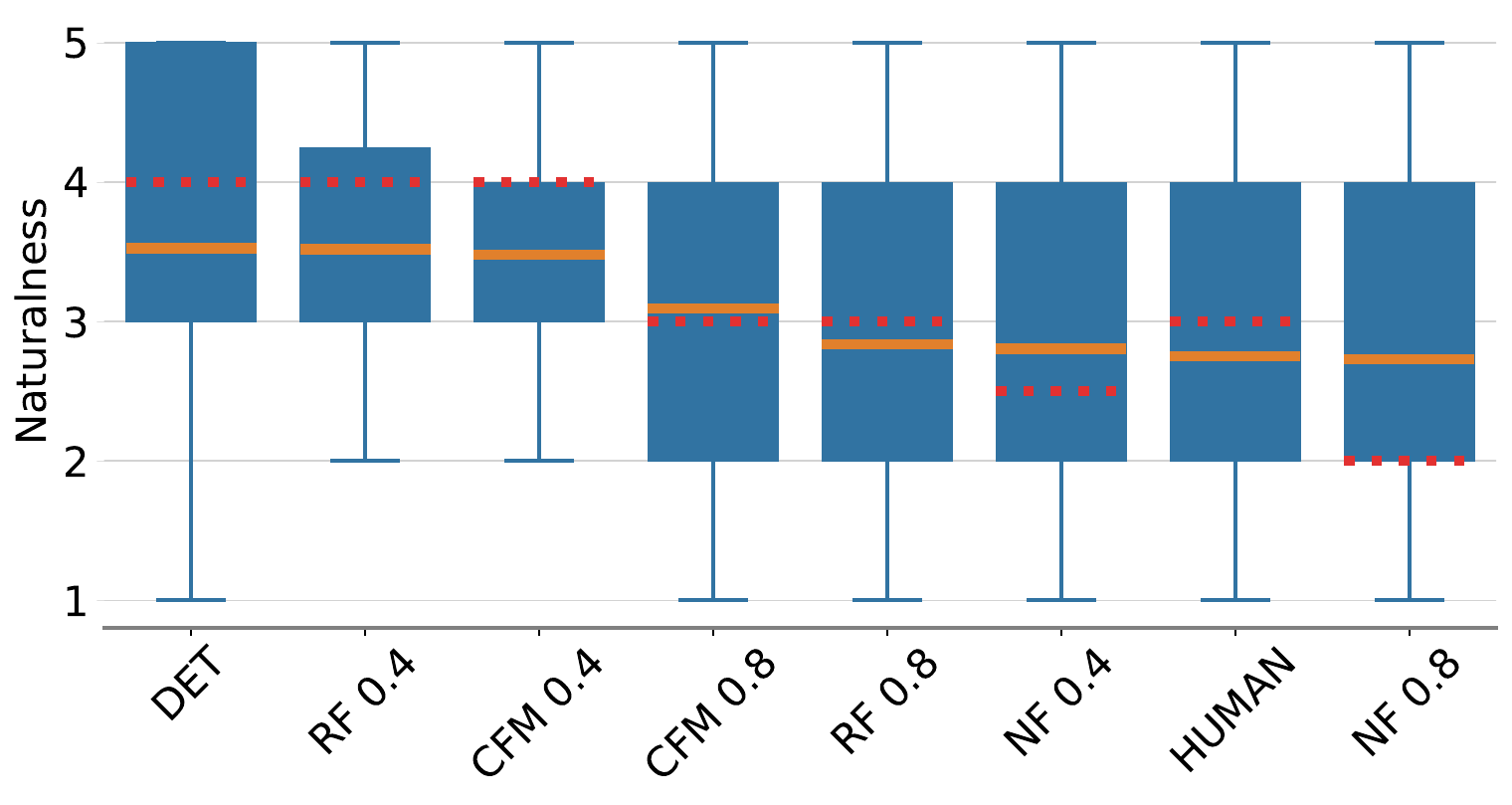}
    \caption{Ratings for the question: Given three samples, how natural does the intonation sound overall?}
    \label{fig:nat}
\end{figure}

\begin{figure}[t]
    \centering
    \includegraphics[width=\linewidth]{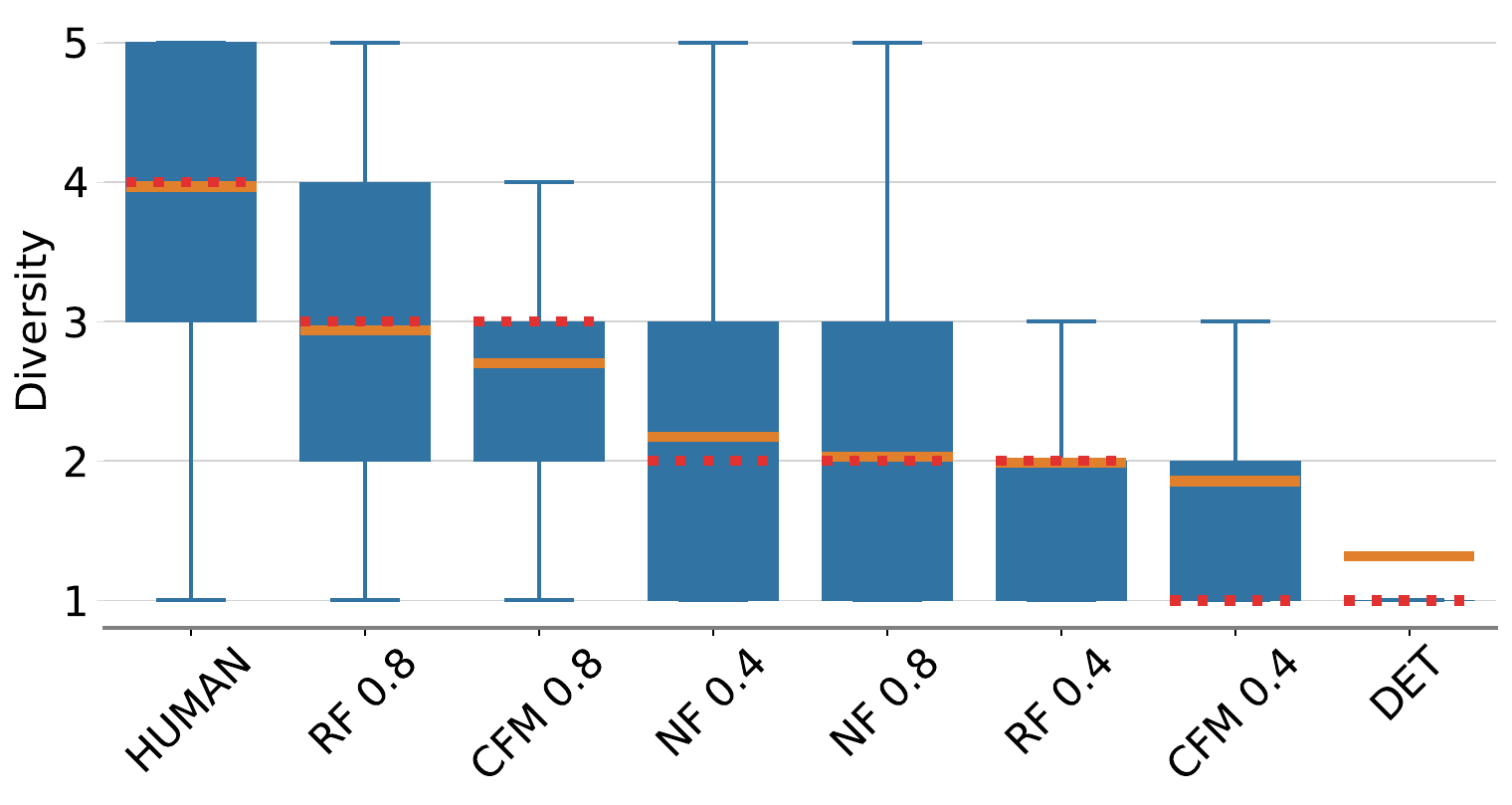}
    \caption{Ratings for the question: Across three samples, how distinct do the intonations sound?}
    \label{fig:var}
\end{figure}
Since both the naturalness of prosody, as well as the perceived diversity across multiple takes of the same sentence by a speaker are difficult to quantify objectively, we additionally conduct a subjective evaluation with human evaluators. The participants of this study were recruited from the crowd-sourcing platform Prolific. They self-identified as English native speakers and were compensated fairly according to high ethical standards. In this study, we asked 40 participants to listen to sets of three audios generated by the same system. Then, we asked them to rate how natural the prosody of the utterances sounded and how distinctive they perceived the three takes to be, using a 5-point Likert scale. This setup of three audios and two questions was repeated across eight different conditions (Human, Deterministic, NF low/high temperature, CFM low/high temperature, RF low/high temperature), with two speakers per system (one male, one female, randomly chosen from LibriTTS) and two sentences per speaker. We thus collected a total of 7680 ratings from the 40 raters, which is a total of 960 ratings per condition split over two questions. In order to isolate differences in prosody as the only variable we adapt, we followed the prosody cloning procedure outlined in \cite{lux2023exact} to convert the human samples to synthetic samples.
These maintain the exact prosody of the human samples, but do not differ in voice or audio quality. The results of the naturalness evaluation are shown in Figure \ref{fig:nat} and the results for the perceived diversity in Figure \ref{fig:var}. Each box is based on 480 ratings. The suffixes of the systems refer to the sampling temperature used where applicable. The mean is indicated by the solid line, the median by the dotted line.

% Data analysis:
% - Temperature does lead to significant differences in perceived variance, with CFM and RF --> Nice!
% - Temperature leads to significant differences in perceived naturalness in CFM and RF --> Not nice! So it's a trade-off
% - There are significant differences between:
%   - NF and CFM (for both naturalness and variance)
%   - NF and RF (for both naturalness and variance)
% --> is NF significantly worse than human? In that case, we would have shown that NF is not capable enough for this task and the more advanced ODE based methods are actually necessary
% - All systems have significant differences in variance to the deterministic system --> that's more like a sanity check, but it does tell us that the CFM speech decoder did not add variance, proving our claim that all the variance is accounted for in the prosody predictors.
% - CFM0.4 and RF0.4 have significant differences from human in naturalness --> surprising that the models outperform the human on this task, but it's likely because the human references had very high variance. On the contrary, DET is likely only so high in naturalness, because it has no variance, so it is boring, but safe.
% - No 0.8 systems have difference from human for naturalness --> We can sell this as a win: they are capable enough to match the human performance, even at high variance!
% - All systems have significant difference from human in variance, but RF with 0.8 gets the closest

From this experiment, we see that there is a trade-off between how natural users perceived the prosody of the three audios to be, and how diverse. Even the human samples could not reach maximal diversity and naturalness at the same time.
The deterministic system was perceived as most natural but least diverse, while the human baseline was considered most diverse but, interestingly, not the most natural.
In cases where only naturalness should be the focus, we find that RF at 0.4, CFM at 0.4, and the deterministic baseline tie for best performance (no significant difference found via Kruskall-Wallis test \cite{kruskal1952use} and Dunn posthoc test \cite{dunn1964multiple}).
However, in cases where diversity is also important, the RF model at 0.8 provides the best trade-off between naturalness and diversity, since it produces the highest diversity after the human baseline, while also maintaining a naturalness score on par with the human baseline (no significant difference found via Kruskall-Wallis test and Dunn posthoc test).

\section{Conclusion}
This paper presents a thorough comparison between different stochastic methods for pitch, energy, and duration modeling against a deterministic and a human baseline. Experimental results show that modeling prosodic parameters in a cascading manner yields better performance than modeling them jointly, but different ordering has no significant impact. Subjective evaluations reveal how prosodic diversity across takes and the prosodic naturalness are inversely correlated. We demonstrate how the sampling temperature allows for effective control over the trade-off between these two aspects at inference time. Finally, we conclude that Rectified Flows offer the best overall performance for prosodic modeling compared to alternative state-of-the-art generative approaches.

\FloatBarrier

%\section{Acknowledgments}
% Acknowledgements are not to be included in the review version.

\bibliographystyle{IEEEtran}
\bibliography{bibliography}

\end{document}